# Ab Initio Local Density Approximation Description of the Electronic Properties of zb-CdS

[1]E. C. Ekuma, L. Franklin, G. L. Zhao, J. T. Wang, and D. Bagayoko

Department of Physics

Southern University and A&M College

Baton Rouge, Louisiana 70813, USA

Self-consistent ab- initio electronic energy band structure of zinc blende CdS are reported within the local density functional (LDF) formalism. Our first principle, non-relativistic and ground state calculations employed a local density functional approximation (LDFA) potential and the linear combination of atomic orbitals (LCAO). Within the framework of the Bagayoko, Zhao, and Williams (BZW) method, we solved self-consistently both the Kohn-Sham equation and the equation giving the ground state density in terms of the wave functions of the occupied states. Our calculated, direct band gap of 2.39 eV, at the $\Gamma$ point, is in accord with experiment. Our calculation reproduced the peaks in the conduction and valence bands density of states, within experimental uncertainties. So are the electron effective mass.

---

[1] Present Address: Department of Physics and Astronomy, Louisiana State University, Baton Rouge, LA 70803

# I. Introduction and Motivation

Cadmium sulfide has long been the most studied of II-VI compounds of the chalcogenide family [1]. Its electronic properties have attracted intensive studies over the past three decades, owing to its great, potential applications. CdS is utilized in devices suitable for optoelectronics [2], piezo-electronics [3], thin film hetero-junction solar cells and semiconducting materials [3]. CdS has been used in making biological labels [4] and quantum-dot lasers [5]. CdS has a commercial use as a phosphor, in photovoltaic cell, a field effect transistor, a hetero-junction laser, beam splitters, photo-resistors and acoustic amplifiers.

The theoretical studies of CdS have had difficulty in predicting some important parameters that generally could have improved the fabrication of devices based on CdS. For instance band structure calculations have had difficulty in locating precisely the Cd *d* state which is experimentally some 8-10 eV [6,7] below the Fermi energy. Also are the electronic band gaps and other related electronic properties.

One of the earliest theoretical studies of the band structure of CdS (zb-CdS) was the local combination of atomic orbitals (LCAO) of Zunger and Freeman [8]. These authors found a room temperature band gap of 2.62 eV and valence band width of 2.75 eV. Their self-consistent exchange correlation model at lattice constant of 5.818 Å gave a zero temperature band gap of 2.01 eV. Stukel et al. [9], employing the self-consistent orthogonalized plane wave (SCOPW) method, found a band gap of 2.72 eV, with the position of the Cd-*4d* state at -16.6 eV. They also calculated empirical-OPW band gap of 2.5 eV, with the position of the Cd-*4d* state at -5.7 eV.

From the mid 1990s to present, several other theoretical reports followed the ones above. The LDA result of Hussain [10] employing FP-LAPW found a band gap of 1.45 eV for the cubic zinc-blende CdS. In their quasi-particle band structure calculation for zb-CdS using a fitting process, Rohlfing et al. [11] found band gaps of 3.70, 1.50, 1.60, 2.45 eV at calculated lattice constants of 5.05, 5.61, 5.72, 5.73 Å respectively. Their corresponding LDA results in the same order of lattice constants were 2.16, 0.78, 0.86, and 0.83 eV.

Indeed, while the LDA calculations of Zakharov et al. [12] led to a band gap of 1.37 eV for zb-CdS, their quasi-particle calculations reported 2.83 eV.

In contrast to the wide range covered by the electronic band gap of zb-CdS as calculated by theory, basic electronic properties of zb-CdS have been experimentally established using different measurement techniques [3,13-16]. Photoluminescence measurements of Lozada-Morales et al. [16] found a band gap of 2.4 eV for zb-CdS. Davila-Pintle et al. [17], in their characterization of films developed by chemical bath technique, found a band gap of 2.42 eV. Ates et al. [18], in their CdS films prepared by successive ionic layer adsorption and reaction (SILAR) method found a band gap of 2.42 eV. The spectroscopic ellipsometry study of zb-CdS by Rossow et al. [19], reported a band gap of 2.40 eV and the photoluminescence measurement of Yu et al. [20], led to a band gap of 2.37 for zb-CdS. The Auger electron spectroscopy of Wilke et al. [21] reported a band gap of 2.42 eV.

From the foregoing, it can be seen that previous electronic structure calculations for zinc blende cadmium sulfide have failed to reproduce the experimental value of the band gap and other related properties. The calculated electronic band gap and other related properties of zinc-blende cadmium sulfide cover a wide range of values. The theoretically calculated band gaps range from 1.58 (i.e., 1.14 to 2.72 eV) (for LDA) and 0.86 eV to 3.70 eV (for GGA, EXX, GW). The experimental band gaps range from 2.37 eV to 2.50 eV, depending on measuring temperature, with the highest value at very low temperature.

While the above failure of previous calculations is a key motivation for this work, its purpose includes placing theoretical computation in a position for informing correctly and guiding practical design and fabrication of devices based on CdS. The shallow *d* bands and their higher ionicity make theoretical description quite difficult. The intrinsic and extrinsic states around the band gap (2.37-2.50 eV) have been extensively studied, but the basic questions about its extended band structure remain unanswered.

This failure means that theory could not adequately guide in the design and fabrication of semiconductor; hence, the very expensive, time-consuming, experimental trials and errors approach had to continue to burden industries. The proliferation of schemes purporting to solve the problems, currently resembles that of the epicycles for the Ptolemaic model of the "Earth" system [22]; the different results of these schemes, most of which entail additional computational parameters unrelated to DFT further complicate the search for first principle solutions. In particular, it clouds the search for the actual limitations of DFT and of the schemes purporting to correct or to go entirely beyond DFT in studying semiconductor materials.

## II. Method

Our calculations employed the local density functional potential of Ceperley and Alder [23], which they obtained by quantum Monte Carlo technique as parametrized by Vosko, Wilk, and Nusair [24] and the linear combination of atomic orbitals. The radial parts of these orbitals are Gaussian functions. We utilized a program package developed at Department of Energy, Ames Laboratory, Iowa [25]. Our calculations are non-relativistic, ground state, *ab-initio* and are performed at zero temperature ($0^oK$). The distinctive feature of our approach resides in our implementation of the Bagayoko, Zhao, and Williams (BZW) method consisting of concomitantly solving self-consistently two coupled equations. One of these equations is the Schrödinger type equation of Kohn and Sham [26], referred to as the Kohn-

Sham (KS) equation. The second equation, which can be thought of as a constraint on the KS equation, is the one giving the ground state charge density in terms of the wave functions of the occupied states. The essentials of the BZW method has been rigorously discussed in literature [22,28-30], but the essentials involves carrying out several self consistency calculations in search of the optimal basis set, contrary to the use of arbitrary chosen basis set, which may not be complete or unduly over-complete.

In search of the BZW optimal basis set, the rigorous to replicate our results follows. Zinc Blende cadmium sulfide (zb-CdS) is a member of the II-VI family, possessing a face centered cubic structure in the space group $T_2^d - F\bar{4}3m$, with space group number 216 and Patterson symmetry $Fm\bar{3}m$ [31]. The zb-CdS unit cell contains two atoms: one cation and one anion at positions as indicated between parentheses: $Cd:(0,0,0)$, $S:(1/4,1/4,1/4)$ [32].

We carried out six (6) different self-consistent calculations utilizing six (6) different basis sets. Table I contains the basis sets utilized for the different six (6) self-consistent calculations. Methodical increase of the basis set shows that calculation V is the one with the optimal basis set. Hence, the electronic structure and related properties presented here were obtained with basis set V, the optimal basis set.

Our self-consistent computations were performed at the experimental lattice constants of 5.82 Å [32,33]. In order to know the exact charge transfer between *Cd* and *S*, we initially, carried out a neutral charge calculation ($Cd^0$ and $S^0$) and found their charges to be +2 for Cd and -2 for S. We then, performed *ab initio* calculation for $Cd^{2+}$ and $S^{2-}$. The wavefunctions from these calculations were inputs for the solid and band structure calculations.

A mesh of 28 *k* points, with proper weights in the irreducible Brillouin zone, was employed in the self-consistent (atomic calculations) iterations. A total of 161 weighted k-points were used in the self-consistent (solid) calculations and a total of 152 weighted k-points employed to generate the energy eigenvalues for the electronic density of states, and the k points were chosen along the high symmetry points in the Brillouin zone of zb-CdS. The electron mass and other related electronic properties were calculated and compared to experimental values.

The high symmetry points utilized in the Brillouin zone are: Σ, Γ and Λ, with the positions for the high symmetry points as: $\Sigma = \frac{\pi}{a}\left(\frac{1}{2},\frac{1}{2},\frac{1}{2}\right)$, $\Gamma = (0,0,0)$ and $\Lambda = \frac{2\pi}{a}(1,0,0)$. The computational error for the valence charge was about 0.0002177 for 44 electrons, a little greater than $10^{-6}$ per electron. The self-consistent potentials converged to a difference around $10^{-5}$ after about 60 iterations.

### III. Results

A total of six (6) different basis sets for zb-CdS were employed in the search of the optimal basis set. These calculations are merely intended to show that a single trial basis set does not generally lead to a correct DFT description of the properties of non-metallic materials. Our calculated, *ab-initio*, self-consistent bands for zb-CdS (Calculation V) are as shown in Fig. 1. As per the comparison with experiment, these bands reproduced most experimental results both in the valence and conduction band. The calculated width of the group of upper valence *p* bands is 4.31 eV, in basic agreement with the experimentally reported value of 4.2 eV [34]. The reported, experimental width [34] of the low laying cadmium 4d valence bands of $1.6 \pm 0.1$ eV is some-what larger than the calculated value of 1.3 eV. Experiment [6] place these 4d bands between 8-10 eV, our calculated minimum of these bands, at Gamma, is 8.39 eV. We also calculated the minimum of the Cd-*4d* bands at the Γ point to be -12.17 eV. Table II shows a detailed comparison between some measured valence band eigenvalues [34] and results from our work.

The comparison plot of the electron energy bands for basis set V (solid lines) and basis set VI (dashed lines) is shown in Fig. 2. As can be seen, the eigenvalues of the occupied states totally converged within computational error, showing clearly that Calculation V is that with the optimal basis set.

Figs. 3 and 4 show the calculated, total (DOS) and partial (pDOS) density of states. We found a peak at $-1.61\pm0.1$ eV in the valence bands DOS. It is in excellent agreement with the experimental value of -1.5 measured by et al [34].

We predict two other peaks at $-2.3\pm0.1$ eV and -3.80 eV for which we could not find an experimental value. Our work located peaks for the conduction bands at $5.1 \pm 0.1$ eV, $6.5\pm0.1$ eV, and 8.6 eV. Note that the observed very small differences between our calculated eigenvalue energies in the DOS and bands, and that measured by Stampfl et al [33]. may be due to the low electron concentration of their samples which ($(5-7)\times10^{17} cm^{-3}$, they used sample thickness of 100 Å) are known to be a major determinant of electronic properties of materials [30]. The calculated electron effective masses in the Γ-Σ, Γ–Λ, and Γ-Δ directions are 0.24, 0.27, and 0.27 $m_o$, respectively. The effective mass is a measure of the curvature of the calculated bands. The agreement between calculated and measured electronic effective masses indicates an accurate determination of the shape of the bands. We thus, find an excellent agreement between our calculated values and measured ones.

### IV. Discussion

Our calculated electronic properties of zb-CdS provide far better agreement with experiment than any other non-BZW theoretical calculations known to us. The above agreements between our LDA-BZW results and experimental ones indicate the accurate density functional theory description of zb-CdS, provided that one looks for and obtains an optimal basis set that is verifiably complete for the description of the *ground state*, on the one hand, and that is not unduly large on the other hand.

The pseudopotential calculation of Vogel et al. [35] placed the *d*-band at -6.8 eV, -10.5 eV, -9.7 eV with LDA, self-interaction corrected, and self-interaction corrected with relaxation calculations respectively. Our calculated position of the *d*-band below the Fermi level is 8.39 eV in basic agreement with experimental values of range 8-10 eV.

The partial density of states gives the true picture of the various orbitals used in the formation of the bands. From Fig. 3, it can be seen that both the valence and conduction bands are formed basically by the hybridization between the various cadmium orbitals with almost no contribution from the sulfur atom. This shows that the density of states for zb-CdS is formed by complex cationic states. The other non-BZW theoretical calculations reported energy band gaps in the range of 0.78 eV to as high as 3.70 eV. Our calculated band gap is 2.39 eV in excellent agreement with experiment.

## V. Conclusion

We have performed a first principle computational study of the electronic and related properties of zb-CdS within density functional theory (DFT). We utilized the local combination of atomic orbitals (LCAO) as implemented within the Bagayoko Zhao Williams (BZW) formalism to avoid the spurious effect associated with basis sets in calculations involving the variational method of the Rayleigh-Ritz type. The electronic band structures, effective masses, total and partial densities of states have been calculated from the self-consistent potentials. The almost perfect agreement between our results and experimental data further underscores the high quality of our method.

Our ab-initio, self-consistent, non-relativistic, ground state LDA-BZW calculations led to electronic and related properties that mostly agree with experiment. Specifically, the calculated band gap of 2.39 eV is in accord with experiment. Our calculations reproduced measured peaks in the conduction band density of state. These agreements point to the accuracy of the density functional description of zb-CdS, provided one utilizes a basis set that is complete for the description of the ground state and that is not over-complete.

The need for the BZW method in self-consistent calculations of electronic properties follows from the Rayleigh theorem, particularly for materials where an energy or band gap exists between occupied and empty states. With the method, the prospects seem great for DFT to inform and to guide the design and fabrication of semiconductor based devices. Further, theory could aid in the search for novel materials with desired properties, including binary to quaternary systems.

**Acknowledgments**: This work was funded in part by the Louisiana Optical Network Initiative (LONI, Award No. 2-10915), the Department of the Navy, Office of Naval Research (ONR, Award Nos. N00014-98-1-0748 and N00014-04-1-0587), the National Science Foundation (Award No. 0754821), and Ebonyi State, Federal Republic of Nigeria (Award No: EBSG/SSB/FSA/040/VOL. VIII/039).

**Table I.** Search for the optimal basis sets (Orbital added is in bold), as per the BZW method, for the description of the valence states of zinc blende cadmium sulfide (zb-CdS). The optimal basis set is that from Calculation V.

| Basis Set | Cadmium Core | Sulfur Core | Cadmium Valence | Sulfur Valence | Total No. of Valence orbitals | Band Gap (eV) |
|---|---|---|---|---|---|---|
| I | 1s2s2p3s3p | 1s | 3d4s4p4d5s | 2s2p3s3p | 23 | 3.06812 |
| II | 1s2s2p3s3p | 1s | 3d4s4p4d5s**5p** | 2s2p3s3p | 26 | 3.31029 |
| III | 1s2s2p3s3p | 1s | 3d4s4p4d5s5p**6s** | 2s2p3s3p | 27 | 2.62036 |
| IV | 1s2s2p3s3p | 1s | 3d4s4p4d5s5p6s**5d** | 2s2p3s3p | 32 | 2.53041 |
| *V* | *1s2s2p3s3p* | *1s* | *3d4s4p4d5s5p6s5d* | *2s2p3s3p**4p*** | *35* | *2.39026* |
| VI | 1s2s2p3s3p | 1s | 3d4s4p4d5s5p6s5d | 2s2p3s3p4p**4s** | 36 | 1.78835 |

**Table II.** Selected, calculated valence band eigenvalues [E(k) in eV] of zb-CdS as compared to experimental data of Stampfl et al. [34]

| Symmetry Label | Level | Experiment E(k) | Calculated E(k) | Symmetry Label | Level | Experiment E(k) | Calculated E(k) |
|---|---|---|---|---|---|---|---|
| $\Gamma$ | $\Gamma_{15}$ | 0.00 | 0.00 | W | $W_1$ | -3.76 | -3.77 |
| | $\Gamma_{12}(4d)$ | -9.00 | -7.92 | | W (4d) | -8.9 | -8.08 |
| | $\Gamma_{12}(4d)$ | -9.50 | -8.39 | | W (4d) | -9.3 | -8.16 |
| X | $X_5$ | -1.54 | -1.88 | K | $K_3$ | ?? | -1.52 |
| | $X_3$ | -3.70 | -3.80 | | $K_1$ | -3.76 | -3.73 |
| | $X_{12}(4d)$ | -8.86 | -8.06 | | K (4d) | -8.90 | -8.12 |
| | $X_{12}(4d)$ | -9.30 | -8.15 | | K (4d) | -9.30 | -8.20 |
| | $X_1$ | -13.85 | -11.70 | | | | |
| L | L3 | -0.62 | -0.73 | | | | |
| | $L_1$ | -4.20 | -4.31 | | | | |
| | L (4d) | -9.25 | -7.89 | | | | |
| | L (4d) | -9.85 | -8.27 | | | | |

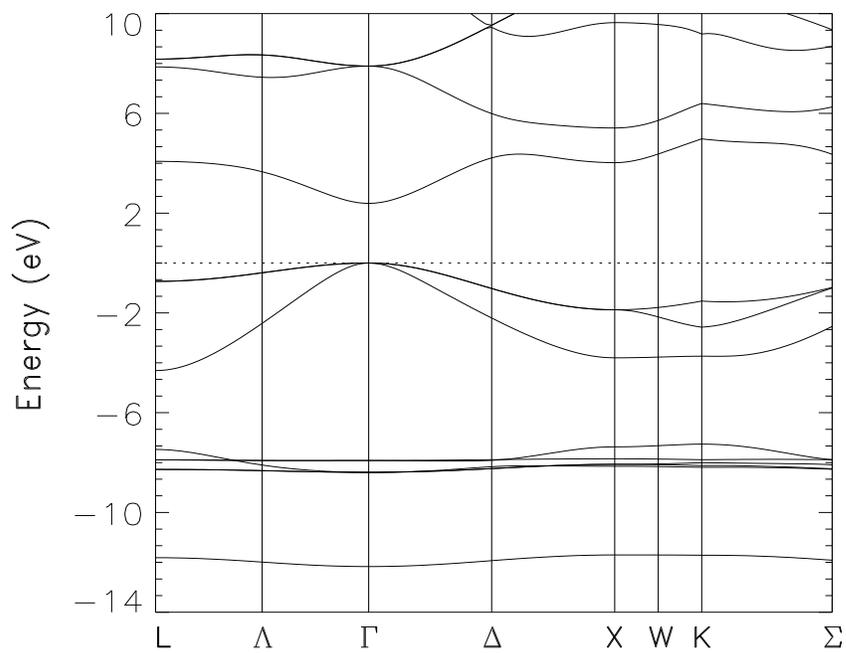

**Fig. 1.** The calculated electronic energy bands of zb-CdS, from Calculation V.

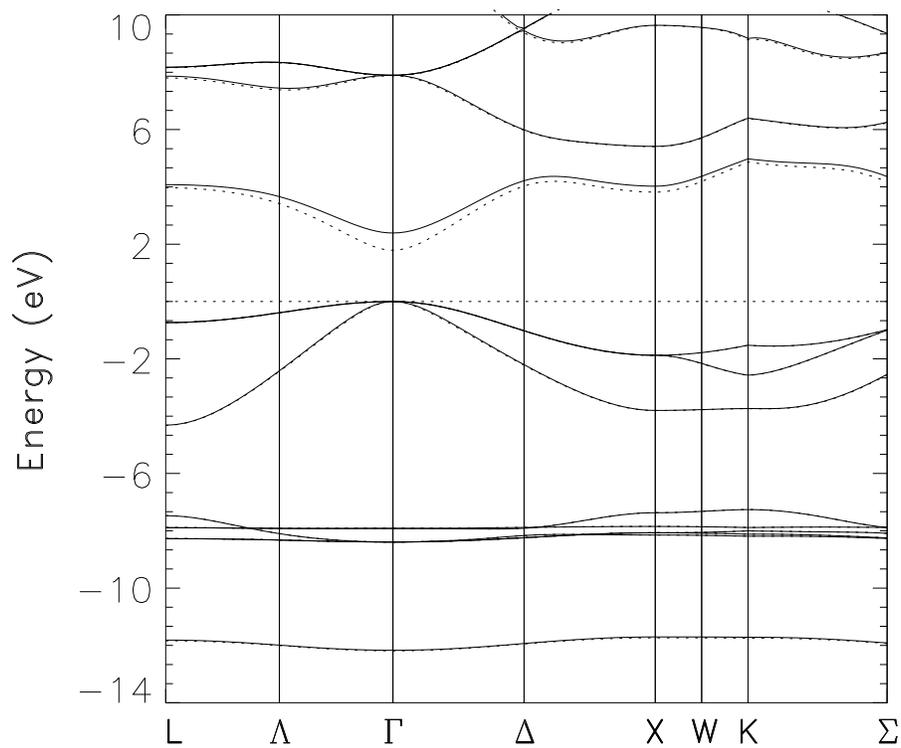

**Fig. 2.** The calculated electronic energy bands of zb-CdS from Calculations V (full lines) and VI (dashed lines). The occupied energies from Calculations V and VI are equal for any *k*-point.

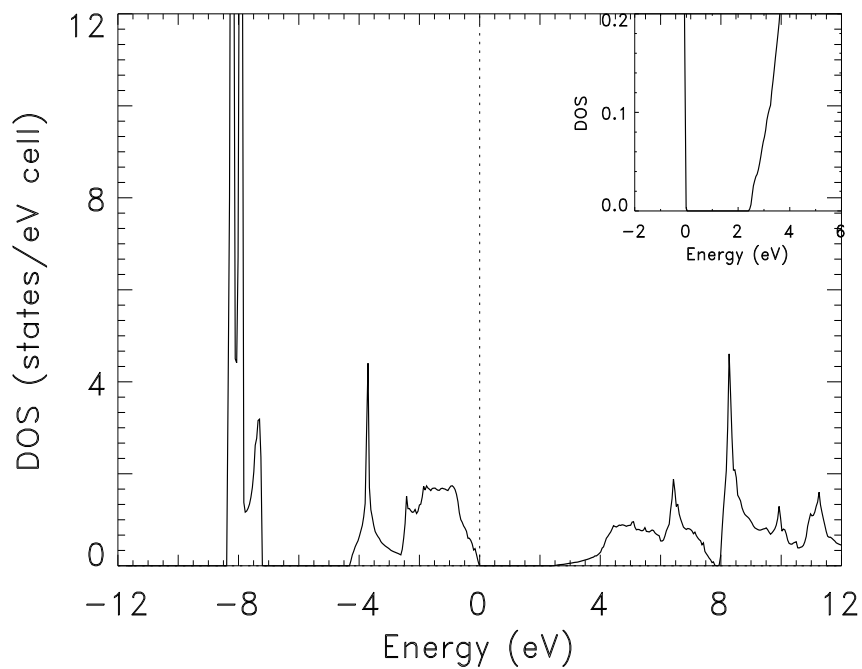

**Fig. 3.** The calculated density of states (DOS) of zb-CdS, obtained with the BZW optimal basis set (i.e. Calculation V).

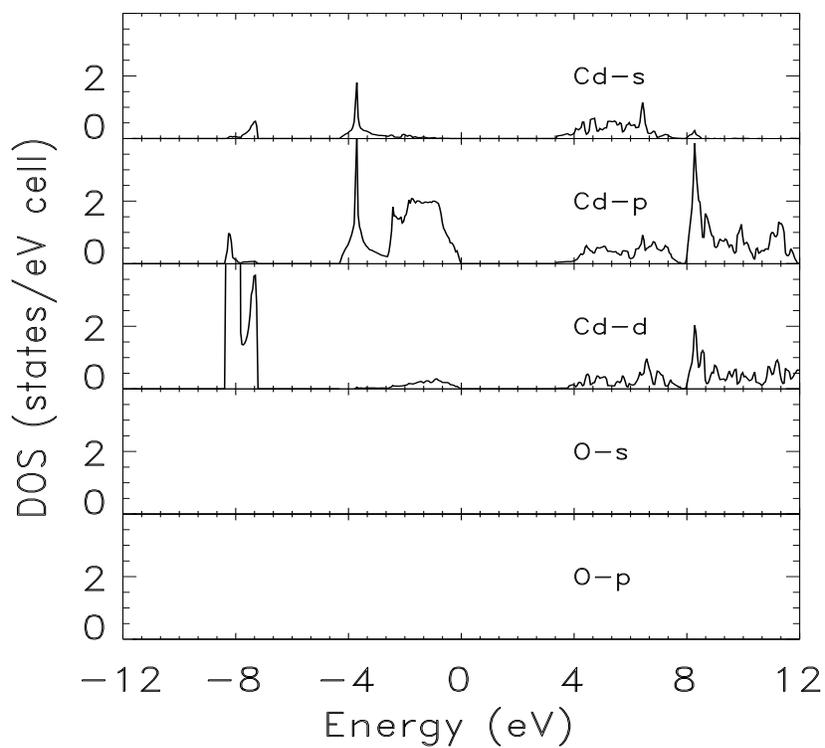

**Fig. 14.** The calculated partial density of states (pDOS) of zb-CdS, obtained with the BZW optimal basis set (i.e. Calculation V).